\documentclass[aps,twocolumn,floatfix,preprintnumbers]{revtex4}
\begin{document}
\preprint{UH-511-1046-04}
\title{
Radiative mass generation and
suppression of supersymmetric contributions to
flavor changing processes}
\author{Javier Ferrandis}
\email{ferrandis@mac.com}
\homepage{http://homepage.mac.com/ferrandis}
\affiliation{Department of Physics \& Astronomy\\
 University of Hawaii at Manoa\\
 2505 Correa Road\\
 Honolulu, HI, 96822}
\begin{abstract}
We explore the possibility that the masses for the first two generations of fermions
and the quark flavor violation are generated radiatively 
in the Minimal Supersymmetric Standard Model. We assume that  
the source of all flavor violation resides in the
the supersymmetry breaking sector and 
is transmitted radiatively to the Standard 
Model fermion sector through finite corrections at low energy. 
The approximate radiative alignment between 
the Yukawa and soft supersymmetry breaking matrices, 
helps to suppress some of the supersymmetric contributions
to flavor changing processes, 
overcoming current experimental constraints.
This mechanism may also explain the non-observation of proton decay,
since flavor conservation in the superpotential
would imply the suppression of dimension five operators
in supersymmetric grand unified theories.
\end{abstract}
\maketitle
%
\section{Introduction}
An outstanding unsolved problem in the Standard Model is
the origin of the fermion mass hierarchies.
A related puzzle is the origin of the flavor
violation observed in the quark sector.
These two problems appear to be connected, since the 
fermion mass hierarchies cannot be explained with precision 
without a theory of flavor.

Many models have been proposed 
to explain the fermion mass hierarchies and the quark mixing angles.
Some of the most popular are
mass-matrix texture models, 
higher-order non-renormalizable operators,
horizontal symmetries, and fixed point mechanisms.
While some of the proposed theories can fit some of the
experimental data they lack a convincing and predictive 
basic principle to explain the origin of the fermion hierarchies.
Fixed point models, in particular, have provided an explanation for 
the top quark mass \cite{Ferrandis:2002ws}
but lack predictivity in a realistic three-generation model.

There is another possibility, which has not received much attention 
lately. The fermion mass hierarchies 
suggest that the masses of the lighter fermions 
may arise only as higher-order radiative effect.
Following the original suggestion by S.~Weinberg \cite{Weinberg:1971nd,Weinberg:1972ws} 
of a mechanism to generate the electron mass radiatively 
from a tree-level muon mass,
several proposals in the framework of non-supersymmetric models were published.
The program was considered more difficult to implement 
in the context of supersymmetric ({\it hereafter} SUSY) models, since, 
as pointed out by L.~Iba\~nez, if supersymmetry is spontaneously broken
only tiny fermion masses can be generated radiatively \cite{Ibanez:1982xg}.
A few ideas to alleviate this problem have been proposed. 
Especially interesting is the possibility,
originally suggested by 
W.~Buchmuller and D.~Wyler \cite{wyler} and later rediscovered 
in Refs.~\cite{Hall:1985dx,Banks:1987iu,Kagan:1987wf,Ma:1988fp},
that the presence of soft supersymmetry breaking 
terms allows for the radiative generation of quark and charged 
lepton masses through sfermion--gaugino loops.
The gaugino mass contributes the violation of fermionic chirality required by a fermion
mass, while the soft breaking terms provide the necessary 
violation of chiral flavor symmetry.
(Additional implications of this possibility have been
studied in Refs.~\cite{softRadSusy,Borzumati:1999sp,kagan,Arkani-Hamed:1995fq,su5relations}).

In this paper I will analyze, in the context of the minimal supersymmetric Standard Model
({\it hereafter} MSSM), the possibility that the masses 
for the first two generations of fermions, as well as the observed mixing
in the quark sector, can actually be generated radiatively.
I will use to this end low-energy one-loop
finite SUSY threshold corrections coming from flavor violating mixings
in the soft supersymmetry breaking sector. 
I will study the constraints on the flavor violating soft breaking sector
and on the supersymmetric spectra imposed by the experimental data on masses,
mixings and flavor changing processes.
The basic conditions that one can expect from a unified
supersymmetric theory, which provides 
the MSSM boundary conditions at a higher energy scale,
generating fermion masses radiatively are,
\begin{enumerate}
\item
A symmetry or symmetries of the superpotential guarantee
flavor conservation and precludes tree-level 
masses for the first and second generations 
of fermions in the supersymmetric limit.
\item
The supersymmetry breaking terms receive small corrections, 
which violate the symmetry of the superpotential and are responsible
for the observed flavor physics.
\end{enumerate}
Under the first assumption, one expects the Yukawa matrices
provided as a boundary condition for the MSSM 
at some high energy scale to be of the form, 
\begin{equation}
{\bf Y} = \left[
 \begin{array}{ccc}
 0 & 0 & 0 \\
 0 & 0 & 0\\
 0 & 0 & y
\end{array}
\right], 
\label{Yukansatz}
 \end{equation}
where ${\bf Y}$ stands generically for the $3\times 3$ 
quark and lepton Yukawa matrices.
We observe that the structure of the Yukawa matrices
given by Eq.~\ref{Yukansatz} is independent of the renormalization scale.
Thus, the renormalization group running from the unification scale down
to the electroweak scale cannot generate non-zero entries.
The opposite is not true: flavor violation, if present in the Yukawa matrices,
would transmit to the soft sector through renormalization group running. 
Under the second assumption, 
one expects the soft trilinear matrices generically to look like,
\begin{equation}
{\bf A} = A  {\cal O}(\lambda),
\label{softansatz}
 \end{equation}
where $A$ is a scalar number and 
${\cal O}(\lambda)$ represents generically a dimensionless
flavor violating polynomial matrix
expanded in powers of $\lambda$. 
The flavor violating perturbation parameter, $\lambda$,
is expected to be determined {\it a} {\it posteriori} 
by the ratios between the fermion masses. 
It is known that in an effective field theory format, holomorphic 
trilinear soft supersymmetry-breaking terms 
originate via non-renormalizable operators
that couple to the supersymmetry-breaking
chiral superfields, ${\cal Z}$, and are suppressed 
by powers of the messenger scale $M$. 
These operators would be generically of the form,
\begin{equation}
\frac{1}{M}
\int d^{2}\theta
{\cal Z} H_{\alpha} \phi^{L} \phi^{R} + c.c.,
\label{holoA}
\end{equation}
where ${\cal Z} = {\cal Z}_{s} + {\cal Z}_{a} \theta^{2}$. 
The vacuum expectation value ({\it hereafter} vev)
of the auxiliary component, $\left<{\cal Z}_{a}\right>$,
parametrizes the supersymmetry breaking scale, $M^{2}_{S}$.
If we assume that the vev of the scalar component of ${\cal Z}$ 
vanishes, $\left< {\cal Z}_{s} \right> =0$,
or is much smaller than the messenger scale,  $\left< {\cal Z}_{s} \right> \ll M$,
then no Yukawa couplings would
arise but only trilinear soft breaking terms.
These conditions could be enforced by an O'Raifeartaigh type
model superpotential or 
by continuous or discrete horizontal flavor
R--symmetries.
Flavor violation in the soft terms may arise, for instance, 
if the supersymmetry-breaking sector fields, ${\cal Z}$, transform
non-trivially under flavor symmetries.
In principle, one also expects flavor violating contributions to the soft mass matrices
arising from operators generically of the form,
\begin{equation}
\frac{1}{M^{2}}
\int d^{4}\theta
{\cal Z} {\cal Z}^{\dagger} k^{2} \phi^{(L,R)} \phi^{(L,R) \dagger};
\end{equation}
here flavor indices have been omitted and
$k^{2}$ is a dimensionless parameter determined by the underlying supersymmetric
theory.
The magnitude of flavor violation in the soft mass matrices would 
depend on the particular model of flavor. Henceforth I will assume that the flavor violation
is concentrated in the soft trilinear matrices while the left and right handed
soft mass matrices are flavor conserving,  
\begin{equation}
{\cal M}^{2}_{D_{(L,R)}} = m_{\widetilde{b}}^{2}
\left[
 \begin{array}{ccc}
 \sigma^{2} & 0 & 0 \\
 0 & \sigma^{2} & 0 \\
 0 & 0 & 1 
\end{array}
\right], 
\label{nondegsoftmas}
\end{equation}
where $m_{\widetilde{b}} \simeq k \left<{\cal Z}_{a}\right>/M$.
To add more generality to the analysis
I will also assume that there is squark non-degeneracy in the left and right handed
sectors given by the parameter $\sigma$.
This coefficient measures the non-degeneracy between the sfermion masses
of different generations, $\sigma = \widetilde{m}_{1,2}/\widetilde{m}_{3}$.
A realistic flavor model based on a ${\rm SU(5)}_{V} \times {\rm U(2)}_{H}$ symmetry
which implements the MSSM boundary conditions
as described above \cite{habajavi}.
We must note that the assumption of flavor conservation in the soft mass matrices 
and of approximate degeneracy between first and second generation 
sfermion masses is important to overcome $\Delta m_{K}$ and $\epsilon_{K}$ 
constraints. We will comment with more detail on this remark later.
In the rest of this paper a phenomenological analysis 
of this basic scenario will be implemented,
independently of the particular model that 
provides these MSSM boundary conditions.
I will show that in this scenatio is possible to fit the fermion 
mass ratios, the CKM matrix elements and 
overcome constraints on flavor changing processes. 
\section{Radiative generation of Yukawa couplings}
In the presence of flavor violation in the soft sector, the left and right handed components
of the sfermions mix. For instance, in the gauge basis 
the $6\times 6$ down--type squarks mass matrix is given by, 
\begin{equation}
{\cal M}^{2}_{D} = 
\left[
\begin{array}{cc}
{\cal M}^{2}_{D_{L}} + v^{2}c^{2}_{\beta}{\bf Y}_{D}^{\dagger}{\bf Y}_{D}
 & ({\bf A}_{D}^{\dagger} c_{\beta} - 
\mu {\bf Y}_{D} s_{\beta}) v \\
 ({\bf A}_{D} c_{\beta} - 
 \mu {\bf Y_{D}^{\dagger}} s_{\beta})v & {\cal  M}^{2}_{D_{R}} +
 v^{2}c^{2}_{\beta}{\bf Y}_{D}{\bf Y}_{D}^{\dagger}
 \\
\end{array}
\right],
\end{equation}
where ${\cal M}^{2}_{D_{R}}$ and ${\cal M}^{2}_{D_{L}}$
are the $3 \times 3$ right handed and left handed soft mass matrices
(including D-terms), ${\bf A}_{D}$ is the $3 \times 3$ soft trilinear matrix,
${\bf Y}_{D}$ is the $3 \times 3$ tree-level Yukawa matrix,
$\tan \beta$ is ratio of Higgs expectation values in the MSSM,
$\mu$ is the so-called mu-term (which is allowed in the superpotential) 
and $v= s_{W} m_{W}/\sqrt{2\pi \alpha_{e}}=174.5$~GeV.
${\cal M}^{2}_{D}$ is diagonalized by a $6 \times 6$ unitary matrix, 
${\cal Z}^{D}$.  In general, the dominant finite 
one-loop contribution to the $3 \times 3$ down--type quark Yukawa matrix
is given by the gluino-squark loop,
\begin{equation}
({\bf Y}_{D})_{ab}^{\hbox{rad}} = \frac{\alpha_{s} }{3 \pi} 
m_{\widetilde{g}}^{*} 
\sum_{c} {\cal Z}^{D}_{ac} {\cal Z}_{(b+3)c}^{D*} 
B_{0}(m_{\widetilde{g}}, m_{\widetilde{d}_{c}}), 
\end{equation}
where $\widetilde{d}_{c}$ ($c =1, \cdot \cdot \cdot, 6$)
are mass eigenstates and $m_{\widetilde{g}}$ is the gluino mass.
$B_{0}$ is a known function defined in the appendix.
The radiatively corrected $3 \times 3$ down--type quark
mass matrix is given by,
\begin{equation}
{\bf m}_{D} = v c_{\beta} ( {\bf Y}_{D}+ {\bf Y}_{D}^{\hbox{rad}} ). \\
\end{equation}
If the supersymmetric spectra were degenerate, i.e. $\sigma =1$, one 
would use the mass insertion method to obtain a simple expression
for ${\bf Y}_{D}^{\hbox{rad}}$, 
\begin{equation}
{\bf Y}_{D}^{\hbox{rad}} = \frac{2 \alpha_{s} }{3 \pi} 
m^{*}_{\widetilde{g}}  ({\bf A}_{D} + \mu {\bf Y}_{D} \tan {\beta}) 
F(m_{\widetilde{b}}, m_{\widetilde{b}},m_{\widetilde{g}}),
\end{equation}
where the function $F(x,y,z)$ is given by,
\begin{equation}
F(x,y,z) = \frac{\left[ (x^{2}y^{2} \ln \frac{y^{2}}{x^{2}} + y^{2}z^{2} \ln \frac{z^{2}}{y^{2}}
+ z^{2}x^{2} \ln \frac{x^{2}}{z^{2}} \right]}{(x^{2}-y^{2})(y^{2}-z^{2})(z^{2}-x^{2})} >0.
\end{equation}
We will ignore CP-violating phases from now on.
We note that in the degenerate squark limit, 
if it were not for tree level contribution to the third generation fermion
masses, the radiatively generated Yukawa matrices would be aligned with the
soft trilinear matrices.
In the general case, the way the non--degeneracy in the 
squark spectra affects the predictions is going to depend on 
the flavor structure in the soft trilinear matrices.

To make my case I will pick a simple
texture for the soft trilinear matrix ${\bf A}_{D}$, which is
motivated on a model with 
a horizontal ${\rm U(2)}_{H}$ symmetry \cite{habajavi,Barbieri:1996ww}.
By simple I mean having only one flavor violating parameter \cite{fritzsch},
\begin{equation}
{\bf A}_{D} = A_{b} \left[
 \begin{array}{ccc}
 0 & \lambda^{2}_{d} &  \lambda^{2}_{d}  \\
 \lambda^{2}_{d} & \lambda_{d} & \lambda_{d} \\
  \lambda^{2}_{d}  &  \lambda_{d} & 1 
\end{array}
\right], 
\label{ADcase1}
 \end{equation}
The analysis that will follow can be easily extended to other possible textures.
Let us assume that squark masses of the first
and second generations are degenerate. In this case 
one obtains a simple expression for 
the radiatively corrected down--type quark mass matrix,
\begin{equation}
{\bf m}_{D} =  \widehat{m}_{b}
\left[
 \begin{array}{ccc}
 0 & \gamma_{b} \lambda^{2}_{d} &  \kappa  \gamma_{b} \lambda^{2}_{d} \\
 \gamma_{b} \lambda^{2}_{d} &  \gamma_{b} \lambda_{d} &  \kappa \gamma_{b} \lambda_{d} \\
   \kappa  \gamma_{b} \lambda^{2}_{d}  &   \kappa \gamma_{b} \lambda_{d} & 1
\end{array}
\right],
\label{mDmat}
\end{equation}
where $\kappa$ is a squark non-degeneracy coefficient,
which in the $m_{\widetilde{q}} \geq m_{\widetilde{g}}$ limit 
is given by,
\begin{equation}
\kappa = 
\kappa^{d}_{13} = \kappa^{d}_{31} =
\kappa^{d}_{23} = \kappa^{d}_{32} = 
\sigma^{2} \ln \sigma^{2} /(\sigma^{2} -1).
\end{equation}
In general $\kappa_{ij}^{d}$ is defined by,
\begin{equation}
\kappa^{d}_{ij} = \frac{F(m^{2}_{(\widetilde{d}_{L})_{i}}, m^{2}_{(\widetilde{d}_{R})_{j}},m^{2}_{\widetilde{g}})}{F(m^{2}_{(\widetilde{d}_{L})_{2}}, m^{2}_{(\widetilde{d}_{R})_{2}},m^{2}_{\widetilde{g}})}, 
\label{kappa}
\end{equation}
Here $\sigma$ is the ratio between first/second and third generation 
down--type squarks
introduced in Eq.~\ref{nondegsoftmas},
\begin{equation}
\widehat{m}_{b} = v c_{\beta} \left( y_{b} +  \rho_{D} (1 - \frac{\mu}{A_{b}} y_{b} \tan \beta) \right), 
\end{equation}
and
\begin{equation}
\gamma_{b} =\frac{v c_{\beta}\rho_{D}}{\widehat{m}_{b}}.
\label{gammabEq}
\end{equation}
$\gamma_{b}$ parametrizes the breaking of the alignment 
between soft trilinear and the Yukawa sector caused by the presence of a 
tree-level mass for the bottom quark, 
and $\rho_{D}$ encodes the dependence on the supersymmetric spectra.
For the case $m_{\widetilde{q}} \geq m_{\widetilde{g}}$ and $\sigma \lesssim 2$
one obtains,
\begin{equation}
\rho_{D} =
\frac{2 \alpha_{s} }{3 \pi} \left( \frac{m_{\widetilde{g}}}{m_{\widetilde{b}}} \right)
\left( \frac{{A}_{b}}{m_{\widetilde{b}}}  \right) \left(\frac{1}{\sigma^{2}}\right),
\label{rhoDEq1}
\end{equation}
(for $\sigma >2$ one should substitute $\sigma \rightarrow 2 \ln \sigma$
in the Eq.~\ref{rhoDEq1}).
Although not diagonal in the gauge basis, the matrix ${\bf m}_{D}$ 
can be brought to diagonal form in the mass basis by a biunitary diagonalization,
$ ({\cal V}^{d}_{L})^{\dagger} {\bf m}_{D} {\cal V}^{d}_{R}
=  \left( m_{d}, m_{s}, m_{b} \right)$.
The down--type quark mass matrix given by Eq.~\ref{mDmat}
makes the following predictions for the quark mass ratios,
\begin{eqnarray}
\frac{m_{d}}{m_{s}} &=& \lambda_{d}^{2} ( 1  + 2 \kappa^{2} \gamma_{b} \lambda_{d}
- 2\lambda_{d}^{2} ) + {\cal O}(\lambda_{d}^{4}) 
 , \\ \frac{m_{s}}{m_{b}} &=&  \gamma_{b} \lambda_{d} ( 1  - \kappa^{2} \gamma_{b} \lambda_{d} 
 + \lambda_{d}^{2})
  + {\cal O}(\lambda_{d}^{3}),
 \label{downquarkrats}
 \end{eqnarray}
These formulas approximately reduce to, 
\begin{equation}
\frac{m_{d}}{m_{s}} = \lambda_{d}^{2}, \quad 
 \frac{m_{s}}{m_{b}} =  \gamma_{b} \lambda_{d}. 
\end{equation}
We can relate $\lambda_{d}$ and $\gamma_{b}$
with dimensionless and approximately renormalization scale independent
fermion mass ratios. To first order,
\begin{equation}
\lambda_{d} = \left( \frac{m_{d}}{m_{s}}\right)^{1/2}, \quad 
\gamma_{b} = \left(  \frac{m_{s}^{3}}{m_{b}^{2} m_{d}} \right)^{1/2},
\end{equation}
Using these relations and the running quark masses
determined from experiment (see appendix),
we can determine $\lambda_{d}$ and $\gamma_{b}$. 
In the degenerate squark mass limit, i.e. $\kappa=1$, we obtain,
\begin{eqnarray}
\lambda_{d} &=& 0.209 \pm 0.019, 
\label{Eq:lambdad} \\
\gamma_{b} &=&  0.109 \pm 0.030, 
\label{Eq:gammab}
\end{eqnarray}
We observe that the size of the flavor violating corrections to the
soft breaking terms is determined by quark mass ratios, while 
constraints on the supersymmetric spectra can be derived from the
parameter $\gamma_{b}$. 
One further observe that for this texture 
the squark non-degeneracy affects the 
determination of $\lambda_{d}$ and $\gamma_{b}$ at the next order 
in $\gamma_{b} \lambda_{d}$.
From Eqs.~\ref{gammabEq} \& \ref{rhoDEq1} we obtain the following upper 
bound on the down--type squark non-degeneracy,
\begin{equation}
\sigma =
 \frac{1}{ \gamma_{b}} \left( \frac{v c_{\beta}}{m_{b}}\right)
\left(\frac{2 \alpha_{s}}{3 \pi}\right) \left( \frac{m_{\widetilde{g}}}{m_{\widetilde{d}}} \right) 
\left(\frac{A_{b}}{m_{\widetilde{b}}}\right) \lesssim \frac{30}{t_{\beta}},
\label{sigma}
\end{equation}
where I used $\gamma_{b}=0.1$, $v=174.5$~GeV,   
$\alpha_{s} = 0.117$, $m_{\widetilde{q}} > m_{\widetilde{g}}$ 
and $A_{b} < 2 m_{\widetilde{b}}$. 
This last condition, $A_{b} < 2 m_{\widetilde{b}}$, approximately 
guarantees the stability of the scalar potential.
We conclude from Eq.~\ref{sigma} that it is possible to 
fit the quark mass ratios with an arbitrary squark spectra,
except for large $\tan \beta$ where the condition $m_{\widetilde{q}_{1,2}}  \lesssim 
m_{\widetilde{b}}$ (compatible with a degenerate squark spectra) is required.
In the limit $m_{\widetilde{g}} > 2 m_{\widetilde{q}}$ one obtains,
\begin{equation}
\gamma_{b} = \frac{2 \alpha_{s} }{3 \pi} 
\left( \frac{v c_{\beta}}{ m_{b}}\right) 
\left( \frac{{A}_{b}}{m_{\widetilde{g}}}  \right)  
\ln \left( \frac{{m}_{\widetilde{b}}}{m_{\widetilde{g}}}  \right) 
\lesssim \frac{1.5}{t_{\beta}}.
\label{rhoDEq2}
\end{equation}
This possibility is perfectly viable for  
low $\tan\beta$ and for large $\tan\beta$. For instance,  
if $\tan\beta>25$ one obtains 
$\gamma_{b}<0.05$, which is still compatible with $\gamma_{b}$ 
being a loop factor.

In the up--type quark sector, one can perform a similar analysis. 
Let us assume the following particular texture as a case study,
\begin{equation}
 {\bf A}_{U} = 
A_{t}  \left[
 \begin{array}{ccc}
 \lambda_{u}^{6} & 0 & 0 \\
0 & \lambda_{u}^{2} & - \lambda_{u} \\
 0 & - \lambda_{u} & 1 
\end{array}
\right],
 \end{equation}
which is inspired on a SU(5) unified model with 
a horizontal ${\rm U(2)}_{H}$ horizontal symmetry \cite{habajavi,Barbieri:1996ww}.
Our choice of sign in the entry (23) will be clear later when we calculate
the CKM mixing matrix. 
One can obtain a simple expression for the radiatively corrected up--type quark mass matrix
including squark non--degeneracy,
\begin{equation}
{\bf m}_{U} =  \widehat{m}_{t}
  \left[
 \begin{array}{ccc}
   \gamma_{t} \lambda_{u}^{6} & 0 & 0 \\
  0  &  \gamma_{t} \lambda_{u}^{2} & - \kappa
  \gamma_{t} \lambda_{u} \\
 0 &  -\kappa \gamma_{t}\lambda_{u} &  1
\end{array}
\right],
\label{mUmat}
\end{equation}
where $\kappa =\kappa^{u}_{23}=\kappa^{u}_{32}$ 
is the up--type squark non--degeneracy coefficient, that 
we assume to simplify is the same that as 
in the down-type squark sector,
\begin{equation}
\widehat{m}_{t} = v s_{\beta} \left( y_{t} +  \rho_{U} ( 1 - \frac{\mu}{A_{t}} y_{t} \cot \beta )\right), 
\end{equation}
and
\begin{equation}
\gamma_{t} =\frac{v s_{\beta}\rho_{U}}{\widehat{m}_{t}}.
\label{thetatEq}
\end{equation}
$\rho_{U}$,
in the case $m_{\widetilde{q}} \geq m_{\widetilde{g}}$ 
and $\sigma \lesssim 2$, is given by,
\begin{equation}
\rho_{U} =
\frac{2 \alpha_{s} }{3 \pi} \left( \frac{m_{\widetilde{g}}}{m_{\widetilde{t}}} \right)
\left( \frac{{A}_{t}}{m_{\widetilde{t}}}  \right) \left(\frac{1}{\sigma^{2}}\right).
\label{rhoUEq}
\end{equation}
After diagonalization, 
$ ({\cal V}^{u}_{L})^{\dagger} {\bf m}_{U} {\cal V}^{u}_{R}
=  \left( m_{u}, m_{c}, m_{t} \right)$, one obtains 
the following predictions for the up--type quark mass ratios,
\begin{eqnarray}
\frac{m_{u}}{m_{c}} &=& \lambda_{u}^{4} (1+ \kappa^{2} \gamma_{t} (1 + \gamma_{t}) ) 
   + {\cal O}(\gamma_{t}^{2}\lambda_{u}^{6}) , 
\label{upquarkrats1}\\ 
\frac{m_{c}}{m_{t}} &=&  
\gamma_{t} \lambda_{u}^{2} (1- \kappa^{2} \gamma_{t}) ( 1 -  2 \kappa^{2} 
\gamma_{t}^{2}  \lambda_{u}^{2} )
  + {\cal O}(\lambda_{u}^6).
 \label{upquarkrats2}
 \end{eqnarray}
These approximately reduce to, 
\begin{equation}
\frac{m_{u}}{m_{c}} = \lambda_{u}^{4}, \quad 
 \frac{m_{c}}{m_{t}} =  \gamma_{t} \lambda_{u}^{2}. 
\end{equation}
We can relate $\lambda_{u}$ and $\gamma_{t}$
with dimensionless fermion mass ratios, to first order,
\begin{equation}
\lambda_{u} = \left( \frac{m_{u}}{m_{c}}\right)^{1/4}, \quad 
\gamma_{t} = \left(  \frac{m_{c}^{3}}{m_{t}^{2} m_{u}} \right)^{1/2},
\end{equation}
using the invariant running quark mass ratios 
determined from experiment (see appendix).
In the degenerate limit,
\begin{eqnarray}
\lambda_{u} &=& 0.225 \pm 0.015, 
\label{Eq:lambdau} \\
\gamma_{t} &=&  0.071 \pm 0.018, 
\label{Eq:gammat}
\end{eqnarray}
From Eqs.~\ref{thetatEq} \& \ref{rhoUEq} 
we obtain the following upper 
bound on the up--squarks non-degeneracy,
\begin{equation}
\sigma =
\frac{1}{\gamma_{t}} 
\left( \frac{v s_{\beta}}{\widehat{m}_{t}}\right)
\left( \frac{2 \alpha_{s}}{3 \pi} \right) 
\left( \frac{m_{\widetilde{g}}}{m_{\widetilde{u}}} \right) 
\left( \frac{A_{t}}{m_{\widetilde{t}}} \right) 
\lesssim 0.6.
\end{equation}
This implies that a certain amount of non--degeneracy in the up--type squark
sector seems to be required, $m_{\widetilde{t}} > 1.5 m_{\widetilde{u}}$, 
to account for the observed quark mass ratios.
One may also notice a similarity in the values 
for $\lambda_{d}$ and $\lambda_{u}$
in Eqs.~\ref{Eq:lambdad} \& \ref{Eq:lambdau} 
and $\gamma_{b}$ and $\gamma_{t}$
in Eqs.~\ref{Eq:gammab} \& \ref{Eq:gammat},
which reflects the curious empirical fact,
\begin{eqnarray}
\left( \frac{m_{d}}{m_{s}} \right) &
\approx & \left( \frac{m_{u}}{m_{c}} \right)^{1/2} \\
\left( \frac{m_{s}^{3}}{m_{b}^{2}m_{d}} \right) &
\approx & \left( \frac{m_{c}^{3}}{m_{t}^{2}m_{u}} \right).
\end{eqnarray}
This may be considered as an experimental evidence 
pointing toward a common underlying mechanism 
generating the up and down--type quark mass matrices. 
An alternative and simpler choice of soft trilinear textures
that makes, to leading order in $\lambda$, 
the same predictions for quark mass ratios is the following,
\begin{equation}
 {\bf A}_{D} = 
 A_{b} \left[
 \begin{array}{ccc}
 0 & \lambda^{2}_{d} &  \lambda^{2}_{d}  \\
 \lambda^{2}_{d} & \lambda_{d} & 2 \lambda_{d} \\
  \lambda^{2}_{d}  &  2 \lambda_{d} & 1 
\end{array}
\right], 
\label{ADcase2}
 \end{equation}
together with,
\begin{equation}
 {\bf A}_{U} = 
A_{t}  \left[
 \begin{array}{ccc}
 \lambda_{u}^{6} & 0 & 0 \\
0 & \lambda_{u}^{2} & 0\\
 0 &  0& 1 
\end{array}
\label{AUcase2}
\right].
 \end{equation}
We will see next that this alternative solution makes also
the same prediciont for the CKM matrix
elements to leading order in $\lambda$.
\subsection{Radiatively generated CKM matrix}
Finally, one can
calculate the CKM mixing matrix, $
{\cal V}_{CKM} =   {\cal V}^{u\dagger}_{L} {\cal V}^{d}_{L}$.
We can express the quark Yukawa
diagonalization matrices as a function of the $\gamma_{u,d}$ and $\lambda_{u,d}$ 
parameters. Assuming the textures given by Eqs.~\ref{mDmat} and \ref{mUmat}
I obtain to leading order in $\lambda_{u,d}$,
\begin{equation}
\left| {\cal V}_{CKM}^{{\rm theo}} \right| =
\left[
 \begin{array}{ccc}
1 - \frac{1}{2} \lambda^{2}_{d}  & - \lambda_{d}   & 
 \gamma_{b} \lambda^{2}_{d}  \\
\lambda_{d} &  1 - \frac{1}{2} \left( \lambda_{d}^{2} +
\gamma_{ud}^{2}
\right) 
& 
- \gamma_{ud}
 \\
 \gamma_{t} \lambda_{u} \lambda_{d} 
 &  \gamma_{ud} 
 & 1 - \frac{1}{2} \gamma^{2}_{ud}
\end{array}
\right], 
\end{equation}
where,
\begin{equation}
\gamma_{ud} =
(\gamma_{t} \lambda_{u} + \gamma_{b} \lambda_{d}).
 \end{equation}
We observe that flavor violation in the entries (12), (21) and (13) of
up--type quark sector is poorly determined
from experiment, since $\lambda_{u}$ only affects to second order 
the corresponding entries of the CKM matrix.
Moreover for the texture under consideration  
$\left|V_{td}\right|$ is of the same order than $\left|V_{ub}\right|$
and it will not appear if not for the flavor mixing in the up-type quark sector.
We can write a simple expression for the predicted CKM matrix at second order if
we simplify and assume that 
$\lambda \approx \lambda_{d} \approx \lambda_{u}$ 
and $\gamma \approx \gamma_{t}\approx \gamma_{b}$.
We obtain,
\begin{equation}
\left[
 \begin{array}{ccc}
1 - \frac{1}{2} \lambda^{2}  &  \gamma \left( 1 -\frac{3}{2}  \gamma \lambda \right)   & 
 \gamma \lambda^{2} \left( 1 + \gamma \lambda) \right)  \\
\lambda  \left( 1 - \frac{3}{2} \gamma \lambda \right)  
&  1 - \frac{1}{2}\lambda^{2} \left( 1 + 4 \gamma^{2} \right) 
&  2 \gamma \lambda \left ( 1 +  \gamma \lambda \right)
 \\
 \gamma \lambda^{2} \left( 1 -  \frac{5}{2} \gamma \lambda \right)  
 &  2 \gamma \lambda \left( 1 +  \gamma \lambda \right)
 & 1 - 2 \gamma^{2} \lambda^{2}
\end{array}
\right], 
\end{equation}
Using the experimentally determined values for $\gamma_{b}$, $\gamma_{t}$,
$\lambda_{d}$ and $\lambda_{u}$
in Eqs.~\ref{Eq:lambdad}, \ref{Eq:gammab}, \ref{Eq:lambdau} \& \ref{Eq:gammat}
I obtain the following central theoretical prediction for the CKM matrix,
$\left| {\cal V}_{CKM}^{{\rm theo}} \right|$,
in the squark degenerate limit, $\kappa=1$, 
\begin{equation}
\left[
 \begin{array}{ccc}
0.977 \pm 0.007 & 0.20 \pm 0.03 & 0.0039 \pm 0.0006 \\ 
0.20 \pm 0.03 & 0.976 \pm 0.008 & 0.047 \pm 0.024 \\
0.005 \pm 0.003 & 0.047 \pm 0.024  & 0.9988 \pm 0.0012
\end{array}
\right], 
\end{equation}
which should be compared with the 90~\% C.L. experimental 
compilation, $\left| {\cal V}_{CKM}^{\hbox{exp}} \right|$,
\begin{equation}
\left[
 \begin{array}{ccc}
0.97485 \pm 0.00075 & 0.2225 \pm 0.0035 & 0.00365 \pm 0.00115 \\ 
 0.2225 \pm 0.0035  & 0.9740 \pm 0.0008 & 0.041 \pm 0.003 \\
0.0009 \pm 0.005 & 0.0405 \pm 0.0035 & 0.99915 \pm 0.00015
\end{array}
\right].
\end{equation}
It appears that there is a very good agreement with 
the experimental data on CKM matrix elements 
from the PDG experimental
compilation. 
It is remarkable that such a simple texture can 
be so close to the experimental results.
The scenario can account for
the observed quark mass ratios and mixing angles
in a natural way, i.e. without
adjusting any supersymmetric parameter
or without resorting to a highly non--degenerate squark spectra.
We note that assuming instead the textures given by the
Eqs.~\ref{ADcase2} and \ref{AUcase2} we obtain the same 
prediction for the CKM matrix to leading order in $\lambda$.
We noticed also that the kind of textures here considered, 
{\it i.e.} with common $\gamma$ coefficients of order ~0.09,
consistent with $\gamma$ being a loop factor,  
in all the entries of the quark 
mass matrices except the (33),
have not been considered before.
The search of Yukawa textures has focused in the past
in polynomial matrices in powers of $\lambda$ with 
coefficients of order 1.

I would like to point out that the
constraints on the supersymmetric spectra are very texture-dependent,
which is positive for the testability of this scenario.
From now on I will use 
$\lambda = \lambda_{u} = \lambda_{d}$.
For instance, assuming a different texture for the soft trilinear matrix 
${\bf A}_{D}$,
\begin{equation}
{\bf A}_{D} = A_{b} \left[
 \begin{array}{ccc}
 0 & \lambda^{3} & 0 \\
 \lambda^{3} & \lambda^{2} & \lambda^{2} \\
 0 & \lambda^{2} & 1 
\end{array}
\right], 
\label{ADcase2}
 \end{equation}
the predictions for the down--type quark mass would be,
\begin{eqnarray}
\frac{m_{d}}{m_{s}} &=& \lambda^{2}+ {\cal O}(\lambda^{4}) 
 , \\ \frac{m_{s}}{m_{b}} &=&  \gamma_{b} \lambda^{2}
  + {\cal O}(\lambda^{4}),
 \label{downquarkrats}
 \end{eqnarray}
The determination of $\lambda$ from 
measured fermion masses would not change at leading order in $\lambda$
but for $\gamma_{b}$ one would obtain $\gamma_{b} \simeq 0.44$. This
is a value much larger than the one we obtained for the texture previously considered
and not compatible with $\gamma$ being a loop factor. 
This case would imply a considerable amount of non--degeneracy in the down--type squark
sector, which would make difficult to fit the CKM elements
except for very low $\tan \beta$.
\subsection{Higher order Yukawa couplings}
Yukawa couplings which are not generated at one loop could be generated
at higher orders. For instance, the Yukawa coupling $({\bf Y}_{U})_{13}$
could be generated at two loops through a diagram with gluino and
Higgs exchange and three soft trilinear vertices:
$({\bf A}_{D})_{12}$, $({\bf A}_{D})_{22}$ and $({\bf A}_{U})_{23}$ \cite{radovan}. 
We are interested in an overestimation of this 2-loop Yukawa
coupling. Assuming that all the sparticles in the loop have masses
of the same order, to maximize the loop factor, we obtain
\begin{equation}
({\bf Y}_{U})_{13}^{\rm 2-loop} \simeq \left( \frac{2 \alpha_{s}}{3 \pi} \right)
\left( \frac{1}{4 \pi} \right)^{2} \left( \frac{v}{m_{\widetilde{q}}}\right)^{2}
c^{2}_{\beta} \lambda^{4},
\end{equation}
here $v=175$~GeV. The ratio $v/m_{\widetilde{q}}$, the $c_{\beta}$ factors 
($c_{\beta} =\cos \beta$)  
and the $\lambda$ factors come from the three
soft trilinear vertices. To facilitate the comparison with the one-loop
generated Yukawa couplings we will express this in powers of $\lambda$.
Using that $\lambda \simeq 0.2$
and $\gamma \simeq 0.1$, we obtain,
\begin{equation}
({\bf Y}_{U})_{13}^{\rm 2-loop} \simeq 
\frac{\gamma \lambda^{10}}{\tan^{2}\beta} 
\left( \frac{1~{\rm TeV}}{m_{\widetilde{q}}}\right)^{2}.
\label{YUsup}
\end{equation}
We note that this 2-loop generated Yukawa coupling is
very suppressed when compared with the one-loop generated couplings
and for all practical purposes it can be considered zero.
\section{Suppression of FCNCs processes by radiative alignment}
Overcoming the present experimental constraints on 
supersymmetric contributions to flavor changing neutral current processes (FCNCs) is 
a necessary requirement for the consistency of any supersymmetric model \cite{susyFC}. 
Correlations between radiative mass generation and dipole operator phenomenology
were first pointed out in Ref.~\cite{Kagan:1994qg}.
For calculational purposes
it is convenient to rotate the squarks to the so-called superKM basis, the
basis where gaugino vertices are flavor diagonal ~\cite{masieroFCNC}. 
The soft trilinear matrix ${\bf A_{D}}$ in the superKM basis is given by,
\begin{equation}
 {\bf A}_{D}^{\rm SKM} =
({\cal V}^{d}_{L})^{\dagger} {\bf A}_{D} {\cal V}^{d}_{R}.
\end{equation}
Assuming the soft trilinear texture from the previous section,
Eq.~\ref{ADcase1},  one obtains, at leading order in $\lambda$ and $\gamma_{b}$,
\begin{equation}
\left| {\bf A}_{D}^{\rm SKM}\right| =
 A_{b}
\left[
 \begin{array}{ccc}
\lambda^{3} &   \gamma_{b} \lambda^{5} & \lambda^{4} \\
\gamma_{b}\lambda^{5} &  \lambda &   (\gamma_{b}-1) \lambda  \\
\lambda^{4} &  (\gamma_{b}-1)  \lambda  &  1 + 2\gamma_{b}\lambda^{2}
\end{array}
\right].
\end{equation}
Using for $\lambda_{d}$ and $\gamma_{b}$ 
the values given by Eqs.~\ref{Eq:lambdad} \& \ref{Eq:gammab}
the amount of soft flavor violation required to 
fit quark masses and mixing angles is determined,
\begin{equation}
\left| {\bf A}_{D}^{\rm SKM} \right|
=  A_{b}
\left[
 \begin{array}{ccc}
 9 \times 10^{-3} & 4 \times 10^{-5} &  2 \times 10^{-3} \\
--  & 0.208  & 0.194  \\
-- &  --  & 1.009
\end{array}
\right].
\label{ADpred}
\end{equation}
The entry most constrained experimentally 
in the soft trilinear mass matrix is the entry (12).
Its contribution to the $K_{L}$--$K_{S}$ mass difference
is, following Ref.~\cite{masieroFCNC}, given by,
\begin{eqnarray}
\Delta m_{K}&= &  \frac{\alpha_{s}^{2}}{216 m^{2}_{\widetilde{q}}} \frac{2}{3} m_{K} f^{2}_{K}
 (\delta_{12}^{d})^{2}_{LR}  [
\left( \frac{m_{K}}{m_{s}+m_{d}}\right)^{2} \times 
\nonumber \\ 
& &\left. \left(  268 ~x f(x) + 144 ~g(x)\right) + 84 ~g(x) \right], 
\label{DmKfor}
\end{eqnarray}
where $x= m_{\widetilde{g}}^{2}/m_{\widetilde{q}}^{2}$,  $m_{\widetilde{q}}=
\sqrt{m_{\widetilde{d}_{L}}m_{\widetilde{d}_{R}}}$ is 
an average squark mass, $m_{K} = 497.6$~MeV, $f_{K}= 160$~MeV,
$\alpha_{s}=0.117$ 
and the functions $f(x)$ and $g(x)$ are defined in the appendix. 
$ (\delta_{12}^{d})_{LR}$ is given by,
\begin{equation}
 (\delta_{12}^{d})_{LR} = \frac{ v c_{\beta} ( {\bf A}_{D}^{\rm SKM})_{12}}{m^{2}_{\widetilde{q}}}.
\end{equation}
Using the predicted flavor violation, Eq.~\ref{ADpred},
for the soft trilinear texture under consideration,
\begin{equation}
 (\delta_{12}^{d})_{LR} = 
 7 \times 10^{-6}
  \left(  \frac{A_{b}}{m_{\widetilde{q}}} \right)
 \left(  \frac{1~\hbox{TeV}}{m_{\widetilde{q}}} \right)
 \frac{ 1}{ t_{\beta}}. 
\end{equation}
Assuming $\tan\beta >5$, $A_{b}< 4 m_{\widetilde{q}}$,
$m_{\widetilde{q}}> 400~$GeV and any gluino-squark mass ratio one obtains 
a contribution to  $\Delta m_{K}$,  $\Delta m_{K} < 10^{-16}$~MeV,
which is below the uncertainty of the experimental measurement,
$\Delta m_{K} = (3.490 \pm 0.006)\times 10^{-12}$~MeV \cite{Hagiwara:fs}.

There is a formula for the the $\Delta m_{B}$ mass difference 
similar to Eq.~\ref{DmKfor}
from which one can obtain constraints 
on $(\delta_{13}^{d})_{LR}$.  $(\delta_{13}^{d})_{LR}$ 
for the texture under consideration, Eq.~\ref{ADpred}, is given by,
\begin{equation}
 (\delta_{13}^{d})_{LR} = 
 3.5 \times 10^{-4}
  \left(  \frac{A_{b}}{m_{\widetilde{q}}} \right)
 \left(  \frac{1~\hbox{TeV}}{m_{\widetilde{q}}} \right)
 \frac{ 1}{ t_{\beta}}. 
\end{equation}
Assuming $\tan\beta >5$, $A_{b}< 4 m_{\widetilde{q}}$,
$m_{\widetilde{q}}> 300~$GeV, and any gluino-squark mass ratio one obtains 
a contribution to $\Delta m_{B}$, $\Delta m_{B} < 7 \times 10^{-13}$~MeV,
which is below the uncertainty of the experimental measurement,
$\Delta m_{B} = (3.22 \pm 0.05)\times 10^{-10}$~MeV \cite{Hagiwara:fs}.
Finally, from the measured $b \rightarrow s \gamma$ decay rate
one can obtain limits on $(\delta_{23}^{d})_{LR}$ through the formula~\cite{Bertolini:1986tg},
\begin{equation}
B(b\rightarrow s \gamma) =  
\frac{2 \alpha_{s}^{2} \alpha}{81 \pi^{2} m^{4}_{\widetilde{q}}} m_{b}^{3} \tau_{b}  m^{2}_{\widetilde{g}} 
 (\delta_{23}^{d})^{2}_{LR} M^{2}(x),
 \end{equation}
where $\alpha^{-1}=127.934$, $\tau_{b} = 1.49 \times 10^{-12}$~s
and the function $M(x)$ is defined in the appendix.
For the texture under consideration, $(\delta_{23}^{d})_{LR}$
is given by,
\begin{equation}
 (\delta_{23}^{d})_{LR} = 
 3.7 \times 10^{-2}
  \left(  \frac{A_{b}}{m_{\widetilde{q}}} \right)
 \left(  \frac{1~\hbox{TeV}}{m_{\widetilde{q}}} \right)
 \frac{ 1}{ t_{\beta}}. 
\end{equation}
Assuming a large value of $\tan\beta$, $\tan\beta >40$, 
$A_{b} \simeq m_{\widetilde{q}}$, a gluino lighter than the squark 
$m_{\widetilde{g}} < m_{\widetilde{q}}$
and $m_{\widetilde{q}}> 400~$GeV 
one obtains a contribution to $B(b\rightarrow s \gamma)$, 
$B(b\rightarrow s \gamma)  < 3.4 \times 10^{-5}$,
which is again below the uncertainty of the experimental measurement,
$B(b\rightarrow s \gamma) = (3.3 \pm 0.4)\times 10^{-4}$ \cite{Hagiwara:fs}
whitout requiring a very heavy squark spectra.
\subsection{Contributions from flavor violating soft masses}
We have shown that the radiative generation of 
quark masses and CKM elements does not necessarily implies 
overcoming constraints on flavor changing neutral current processes.
On the other hand, we expect any theory of flavor generate certain
amount of flavor violation or non-degeneracy in the soft mass matrices.
Even whether the soft mass matrices are diagonal in the 
interaction basis they may not be diagonal in the SKM basis
if the diagonal soft masses are not degenerate in the interaction basis.
Since the most constrained entry is the (12) from the measurement
of $\Delta m_{K}$, we will estimate what is the amount of nondegeneracy that 
we can afford between the first and generation of squark masses.
For instance, the left handed down-type squark mass matrix 
in the SKM basis is given by, 
\begin{equation}
 ({\cal M}^{2}_{D_{L}})^{\rm{SKM}} =
({\cal V}^{d}_{L})^{\dagger}{\cal M}^{2}_{D_{L}} {\cal V}^{d}_{L},
\end{equation}
and analogously for the right handed soft mass matrix.
Assuming that ${{\cal M}}^{2}_{D_{L}}$ is diagonal,
{\it} i.e. there is no flavor violation in interaction basis,
and assuming that the non-degeneracy between first and second generation is,
\begin{equation}
\frac{(m^{2}_{\widetilde{d}_{L}} - m^{2}_{\widetilde{s}_{L}})}{m^{2}_{\widetilde{d}_{L}}} = \lambda^{n}.
\end{equation}
We obtain in the SKM basis
\begin{equation}
(\delta^{d}_{LL})_{12} = \lambda^{n+1},
\end{equation}
and analogously for $(\delta^{d}_{RR})_{12}$. 
This would generate a contribution to $\Delta m_{K}$
~\cite{masieroFCNC} given by,
\begin{eqnarray}
\Delta m_{K}&= &  \frac{2 \alpha_{s}^{2}}{648 m^{2}_{\widetilde{q}}}  m_{K} f^{2}_{K}
 \lambda^{2(n+1)}  [120~x f(x) +168 ~g(x) 
\nonumber \\ 
& &\left. 
\left( \frac{m_{K}}{m_{s}+m_{d}}\right)^{2} \left(  384 ~x f(x) - 24 ~g(x)\right)  \right], 
\label{DmKfor2}
\end{eqnarray}
where we have added the LL and RR contributions
assuming they are of the same size.
The parameters in this formula where defined above.
We obtain that for $n=3$
we can avoid the saturation of the experimental measurement for
squark spectra $m_{\widetilde{q}}> 400~$GeV, while
for $n=2$ the squark spectra must be $m_{\widetilde{q}}> 2~$TeV.
This constraint can be milder if the gluino-squark mass ratio
is much larger or smaller than one.

Using known expressions \cite{masieroFCNC,Bertolini:1986tg}
we can also calculate the size of a possible flavor violating
soft mass to $B( b\rightarrow s \gamma)$.
This is in general suppressed when compared with the LR, {\it i.e.}
soft trilinear contribution. For instance for the LL contribution we
obtain, 
\begin{equation}
\frac{1}{6} \left( \frac{m_{b}}{m_{\widetilde{g}}} \right) \frac{(\delta_{12}^{d})_{LL}}{(\delta_{12}^{d})_{LR}} 
\approx  3 \times 10^{-3} t_{\beta} \left(\frac{m_{\widetilde{b}}}{m_{\widetilde{g}}}\right)
\end{equation}
where we assumed that $(\delta_{12}^{d})_{LL} =  \lambda$.
We used that $v$ and $\gamma_{\tau}$ are given by 
$v=175$~GeV and $\gamma_{b}\approx0.1$. 
 Even considering very large $\tan\beta$ values a possible LL contribution 
 is one order of magnitude smaller than the LR contribution.
 Therefore the approximate constraints on the supersymmetric spectra 
calculated above from the LR contribution to $B( b\rightarrow s \gamma)$, 
while ignoring the LL contributions, are still valid.

To sum up, the flavor violation present in the soft trilinear supersymmetry breaking
sector, which is necessary in this scenario to generate quark mixings radiatively,
is not excluded by the present experimental constraints on
FCNCs processes.
The approximate radiative alignment between 
Yukawa matrices and soft trilinear terms helps to
suppress some of the supersymmetric contributions to FCNCs.
On the other hand if the underlying theory of flavor predicts
non--degeneracy between the masses of first and second 
generation of squarks larger than $\lambda^{2}$
the constraints on the squark spectra,
coming from the measurement of $\Delta m_{K}$,
are much stronger than the ones from the LR contributions.
\section{Charged lepton masses}
The electron and muon masses 
could also be generated radiatively 
through one--loop bino--slepton exchange involving
the soft supersymmetry-breaking terms, analogously to the gluino--squark exchange in the
quark sector. This possibility that was first suggested
in Refs.~\cite{wyler,Ma:1988fp}. 
The radiatively generated lepton Yukawa couplings are in this case given by,
\begin{equation}
({\bf Y}_{L})_{ab}^{\hbox{rad}} = \frac{\alpha }{ 2 \pi} 
m_{\widetilde{\gamma}} 
\sum_{c} {\cal Z}^{L}_{ac} {\cal Z}_{(b+3)c}^{L*} 
B_{0}(m_{\widetilde{\gamma}}, m_{\widetilde{l}_{c}}), 
\end{equation}
where $m_{\widetilde{\gamma}}$ is the photino mass, ${\cal Z}^{L}$
is the slepton diagonalization matrix and $m_{\widetilde{l}_{c}}$ 
($c=1,\cdot\cdot\cdot 6$) are slepton mass eigenstates. 

To make my case I will pick a simple 
texture for the soft trilinear matrix ${\bf A}_{L}$, which is
motivated on a SU(5) unified model plus a 
${\rm U(2)}_{H}$ horizontal symmetry \cite{habajavi,Barbieri:1996ww},
\begin{equation}
{\bf A}_{L} = A_{\tau} 
\left[
 \begin{array}{ccc}
 0 & \lambda^{2}_{l} & \lambda^{2}_{l}  \\
  \lambda^{2}_{l} & 3 \lambda_{l} & \lambda_{l} \\
\lambda^{2}_{l}  &  \lambda_{l} & 1 
\end{array}
\right]. 
\label{AL}
 \end{equation}
I assume that first and second generation slepton masses are 
degenerate and I allow non--degeneracy between third 
and first/second generation, as in the squark sector I will
parametrize the non--degeneracy through the coefficient 
$\kappa$ analogous to the one defined by Eq.~\ref{kappa}.
One obtains then a simple expression for 
the radiatively corrected lepton quark mass matrix,
\begin{equation}
{\bf m}_{L} =  \widehat{m}_{\tau}
\left[
 \begin{array}{ccc}
 0 & \gamma_{\tau} \lambda^{2}_{l} & \kappa \gamma_{\tau} \lambda^{2}_{l} \\
 \gamma_{\tau} \lambda^{2}_{l} &  
3 \gamma_{\tau} \lambda_{l} &  \kappa \gamma_{\tau} \lambda_{l} \\
\kappa \gamma_{\tau} \lambda^{2}_{l} &   \kappa \gamma_{\tau} \lambda_{l} & 1
\end{array}
\right],
\label{mLmat}
\end{equation}
In the $m_{\widetilde{l}} \geq m_{\widetilde{B}}$ limit 
one obtains 
$\kappa = \sigma^{2} \ln \sigma^{2} /(\sigma^{2} -1)$.
$\widehat{m_{\tau}}$ is defined by,
\begin{equation}
\widehat{m}_{\tau} = v c_{\beta} \left( y_{\tau} +  \rho_{L} (1 - \frac{\mu}{A_{\tau}} 
y_{\tau} \tan \beta) \right), \quad
\end{equation}
$\gamma_{\tau}$ is defined by an expression similar
to the one for $\gamma_{b,t}$,
\begin{equation}
\gamma_{\tau} =\frac{v c_{\beta}\rho_{L}}{\widehat{m}_{\tau}},
\label{gammatauEq}
\end{equation}
In the case $m_{\widetilde{l}} \geq m_{\widetilde{B}}$ and $\sigma \lesssim 2$
one obtains,
\begin{equation}
\rho_{L} =
\frac{\alpha }{ \pi} \left( \frac{m_{\widetilde{\gamma}}}{m_{\widetilde{\tau}}} \right)
\left( \frac{{A}_{\tau}}{m_{\widetilde{\tau}}}  \right) \left(\frac{1}{\sigma^{2}}\right),
\label{rhoEEq1}
\end{equation}
(for $\sigma >2$ one should substitute $\sigma \rightarrow 2 \ln \sigma$).
Although not diagonal in the gauge basis the matrix ${\bf m}_{L}$ 
can be brought to diagonal form in the mass basis by a biunitary diagonalization,
$ ({\cal V}^{l}_{L})^{\dagger} {\bf m}_{L} {\cal V}^{l}_{R}
=  \left( m_{e}, m_{\mu}, m_{\tau} \right)$.
The lepton mass matrix given by Eq.~\ref{mLmat}
makes the following predictions for the quark mass ratios,
\begin{eqnarray}
\frac{m_{e}}{m_{\mu}} &=& \frac{1}{9} \lambda_{l}^{2} ( 1 + \frac{5}{3} \gamma_{\tau} \lambda_{l}
-   \frac{2}{9} \lambda_{l}^{2} ) + {\cal O}(\lambda_{l}^{4}) 
 , \\ \frac{m_{\mu}}{m_{\tau}} &=&  3 \gamma_{\tau} \lambda_{l} ( 1 - 
 \frac{1}{3} \gamma_{\tau} \lambda_{l} + 
  \frac{1}{9} \lambda_{l}^{2} )
  + {\cal O}(\lambda_{l}^{3}),
 \label{leptonrats}
 \end{eqnarray}
which approximately reduce to, 
\begin{equation}
\frac{m_{e}}{m_{\mu}} = \frac{1}{9} \lambda_{l}^{2}, \quad 
 \frac{m_{\mu}}{m_{\tau}} = 3 \gamma_{\tau} \lambda_{l}. 
\end{equation}
We can express $\lambda_{l}$ and $\gamma_{\tau}$
as a function of 
dimensionless and approximately renormalization scale independent
charged lepton mass ratios, to first order,
\begin{equation}
\lambda_{l} = 3 \left( \frac{m_{e}}{m_{\mu}}\right)^{1/2}, \quad 
\gamma_{\tau} = \frac{1}{9} \left(  \frac{m_{\mu}^{3}}{m_{\tau}^{2} m_{e}} \right)^{1/2},
\end{equation}
using the invariant running lepton mass ratios 
determined from experiment (see appendix).
In the degenerate limit, {\it i.e.} $\kappa=1$,
one obtains, 
\begin{eqnarray}
\lambda_{l} &=& 0.206480 \pm 0.000002, 
\label{Eq:lambdal} \\
\gamma_{\tau} &=&  0.09495\pm 0.00001, 
\label{Eq:gammatau}
\end{eqnarray}
Interestingly these values of $\lambda_{l}$ and $\gamma_{\tau}$ 
are consistent with the values required in the quark sector
for $\lambda_{u,d}$ and $\gamma_{t,b}$ respectively, unveiling
two surprising relations,
\begin{eqnarray}
\lambda & = &
\left( \frac{m_{d}}{m_{s}} \right) ^{1/2}
\approx \left( \frac{m_{u}}{m_{c}} \right)^{1/4} 
\approx 3 \left( \frac{m_{e}}{m_{\mu}} \right)^{1/2}, 
\\
\theta &=& \left( \frac{m_{s}^{3}}{m_{b}^{2}m_{d}} \right) ^{1/2}
\approx \left( \frac{m_{c}^{3}}{m_{t}^{2}m_{u}} \right) ^{1/2}
\approx  \frac{1}{9} \left( \frac{m_{\mu}^{3}}{m_{\tau}^{2}m_{e}} \right)^{1/2}, 
\end{eqnarray}
where $\lambda$ is related to the Cabbibo angle
and $\theta$ is a new parameter. These coincidences indicate that 
$\lambda$ and $\theta$ are parameters directly connected with the
underlying theory of flavor. 
The coincidence of these
mass ratios may be considered experimental evidence 
supporting the consistency of this scenario. 
Using Eq.~\ref{gammatauEq} and assuming that 
the slepton masses are heavier than the bino mass, 
$m_{\widetilde{e}} \geq m_{\widetilde{B}}$,
one obtains the following constraint for the 
non-degeneracy between first and third generations slepton masses, 
\begin{equation}
\sigma =
 \frac{1}{ \gamma_{\tau}} \left( \frac{v c_{\beta}}{ m_{\tau}}\right)
\left(\frac{ \alpha}{ \pi}\right) \left( \frac{m_{\widetilde{B}}}{m_{\widetilde{e}}} \right) 
\left(\frac{A_{\tau}}{m_{\widetilde{\tau}}}\right) \lesssim \frac{5}{t_{\beta}}.
\label{gammal}
\end{equation}
This would imply a considerable amount of non-degeneracy in the slepton sector
between the first/second and third generations, 
$m_{\widetilde{\tau}} \ge  \frac{1}{5} \,t_{\beta}\,m_{\widetilde{e}} $.
On the other hand if the bino mass is heavier than the selectron mass
the required slepton non--degeneracy is reduced.
It would be interesting if an alternative soft trilinear lepton texture 
could be found that keeps the succesfull prediction given by, 
$\lambda_{l} = 3 \left( \frac{m_{e}}{m_{\mu}}\right)^{1/2}$, 
without requiring slepton non-degeneracy. It would also be interesting if
the corrections to the supersymmetry breaking sector could generate
the important slepton non-degeneracy that seems to be required in the 
slepton sector by the texture here examined.
Nevertheless, we would like to mention that 
there is an alternative possibility that could allow us to fit the lepton mass
ratios without resorting to a highly non-degenerate slepton spectra.
\subsection{Non-holomorphic terms}
The most general softly broken supersymmetric lagrangian can contain 
non-holomorphic operators of the form \cite{Borzumati:1999sp},
\begin{equation}
\frac{1}{M^{3}}
{\cal Z}{\cal Z}^{\dagger} 
H_{\alpha}^{\dagger} \phi_{L} \phi_{R},
\end{equation}
where ${\cal Z}$ are the supersymmetry- and flavor-breaking chiral superfields.
These terms are suppressed for a messenger scale well above the supersymmetry breaking
scale. They can therefore be relevant only with a low scale for both flavor and supersymmetry breaking.
When they are relevant,
they would give rise to additional soft trilinear terms 
relevant for the radiative Yukawa generation in the down-type quark and lepton sectors.
For instance, 
in the slepton sector of the soft breaking lagrangian there is an additional contribution,
\begin{equation}
A_{\tau} {\cal H}_{d} L E + A_{\tau}^{\prime} {\cal H}_{u}^{*} L E +\cdot \cdot \cdot 
\end{equation}
The calculation of the radiative masses is then 
similar to the one implemented above except for a 
shift in the soft trilinear terms by an amount,
\begin{equation}
A_{\tau} \rightarrow A_{\tau} + \tan \beta A_{\tau}^{\prime},
\end{equation}
where the $\tan\beta$ factor comes from the $H_{u}^{*}$ term. The non-holomorphic
contribution, $A_{\tau}^{\prime}$, is dominant for large $\tan\beta$. In such a
case non-degeneracy in the slepton sector would not be required to fit 
the lepton masses.
\subsection{Lepton flavor violating processes}
Overcoming the present experimental constraints on 
supersymmetric contributions to lepton flavor changing processes is 
a necessary requirement for the consistency of our scenario \cite{susyFC}. 
As in the squark sector, for calculational purposes,
it is convenient to rotate the sleptons to the basis
where gaugino vertices are flavor diagonal ~\cite{masieroFCNC}. 
The soft trilinear matrix ${\bf A_{L}}$ in the superKM basis is given by,
\begin{equation}
 {\bf A}_{L}^{\rm SKM} =
({\cal V}^{l}_{L})^{\dagger} {\bf A}_{L} {\cal V}^{l}_{R}.
\end{equation}
Assuming the soft trilinear texture from
Eq.~\ref{AL}, 
one obtains, to leading order in $\lambda_{l}$ and $\gamma_{\tau}$,
\begin{equation}
\left| {\bf A}_{L}^{\rm SKM}\right| =
 A_{\tau}
\left[
 \begin{array}{ccc}
\frac{1}{3} \lambda^{3}_{l} &   \frac{2}{3}\gamma_{\tau} \lambda^{3}_{l} & \frac{2}{3}\lambda^{2}_{l} \\
\frac{2}{3}\gamma_{\tau}\lambda^{3}_{l} &   3 \lambda_{l} &   \lambda_{l}  \\
 \frac{2}{3}\lambda^{2}_{l} &  \lambda_{l}  &  (1 + 2\gamma_{\tau}\lambda^{2}_{l})
\end{array}
\right],
\end{equation}
Using for $\lambda_{l}$ and $\gamma_{\tau}$ 
the values given by Eqs.~\ref{Eq:lambdal} \& \ref{Eq:gammatau}
the amount of soft flavor violation is determined,
\begin{equation}
\left| {\bf A}_{L}^{\rm SKM} \right|
=  A_{\tau}
\left[
 \begin{array}{ccc}
 3 10^{-3} & 5.6 \times 10^{-4} &  3 \times 10^{-2} \\
--  & 0.648  & 0.216  \\
-- &  --  & 1.007
\end{array}
\right],
\label{downpred}
\end{equation}
The entry most constrained experimentally 
in the soft trilinear mass matrix is the entry (12).
Its contribution to the $B(\mu \rightarrow e \gamma)$
is, following Ref.~\cite{masieroFCNC}, given by,
\begin{equation}
\Gamma_{\mu \rightarrow e \gamma}=
\frac{B(\mu \rightarrow e \gamma) }{B(\mu \rightarrow e \nu_{\mu} \overline{\nu_{e}} )} =  
\frac{24 \alpha^{3} \pi }{m_{\mu}^{2}  G_{F}^{2} m^{4}_{\widetilde{l}}}  m^{2}_{\widetilde{\gamma}} 
 (\delta_{12}^{l})^{2}_{LR} M^{2}(x)
\label{muegamma}
 \end{equation}
where $\alpha^{-1}=127.934$, $G_{F} = 1.16639 \times 10^{-11}$~MeV${}^{-2}$
and the function $M(x)$ is defined in the appendix.
For the texture under consideration $(\delta_{12}^{l})_{LR}$
is given by,
\begin{equation}
 (\delta_{12}^{l})_{LR} = 
  10^{-4}
  \left(  \frac{A_{\tau}}{m_{\widetilde{l}}} \right)
 \left(  \frac{1~\hbox{TeV}}{m_{\widetilde{l}}} \right)
 \frac{1}{ t_{\beta}}. 
\end{equation}
Assuming a large value of $\tan\beta >50$, 
$A_{b}\simeq  m_{\widetilde{l}}$, a photino mass 
$m_{\widetilde{\gamma}} < m_{\widetilde{l}}$ and
$m_{\widetilde{l}}> 1~$TeV one obtains 
a branching fraction $\Gamma_{\mu \rightarrow e \gamma} < 8 \times 10^{-12}$,
which is still below the experimental limit,
$\Gamma^{\hbox{exp}}_{\mu \rightarrow e \gamma} 
<  1.2 \times 10^{-11}$ \cite{Hagiwara:fs}, 
without requiring a multiTeV slepton spectra.
On the other hand, this indicates that unless the supersymmetric
spectra is above 1 TeV this process could be observed in the
near future.
A possible contribution to $B(\mu 
\rightarrow e \gamma)$ 
from flavor violating soft masses of the order
$(\delta_{12}^{l})_{LL} \approx  \lambda^{3}$ 
would receive a suppression factor 
compared with the contribution from the soft trilinear terms of the form,
\begin{equation}
\frac{1}{6} 
\left( \frac{m_{\mu}}{m_{\widetilde{\gamma}}} \right) \frac{(\delta_{12}^{l})_{LL}}{(\delta_{12}^{l})_{LR}} 
\approx 5 \times 10^{-3} t_{\beta} \left(\frac{m_{\widetilde{l}}}{m_{\widetilde{\gamma}}}\right).
\end{equation}
where we used that
$v$ and $\gamma_{\tau}$ are given by 
$v=175$~GeV and $\gamma_{\tau} =0.95$.
The predictions for $\Gamma_{\tau \rightarrow e \gamma}$ and
$\Gamma_{\tau \rightarrow \mu \gamma}$ can be calculated from 
Eq.~\ref{muegamma} with the substitutions $m_{\mu} \rightarrow m_{\tau}$
and $\delta_{12}^{l} \rightarrow \delta_{13}^{l},\delta_{23}^{l}$ respectively.
For the texture under consideration, using the same parameter space limits indicated
above, I obtain,
$\Gamma_{\tau \rightarrow e \gamma} < 8 \times 10^{-11}$
and $\Gamma_{\tau \rightarrow \mu \gamma} < 4 \times 10^{-10}$, 
these two are far below the present experimental limits,
$\Gamma^{\hbox{exp}}_{\tau \rightarrow e \gamma} 
<  2.7 \times 10^{-6}$ and $\Gamma^{\hbox{exp}}_{\tau \rightarrow \mu \gamma} 
<  1.1 \times 10^{-6}$ \cite{Hagiwara:fs}.
\subsection{Massless neutrinos}
The generation of small neutrino masses required by the experiment is
an open problem for this scenario.
It is not possible to generate radiatively a Majorana or Dirac neutrino Yukawa matrix
through this mechanism in the MSSM, since even if a right-handed neutrino 
is introduced it cannot give rise to a radiatively generated
Dirac Yukawa matrix, because it carries no 
${\rm SU(3)}_{C}\times {\rm SU(2)}_{L}\times {\rm U(1)}_{Y}$ quantum numbers.
The solution to this problem most probably will require
to enlarge the particle content and the symmetries of the model. 
\section{Proton decay supression}
The mechanism of soft radiative generation of Yukawa couplings could be embedded
in the particular case of a supersymmetric grand unified theory.
It is believed that strong experimental limits on proton decay place
stringent constraints on supersymmetric grand unified models.
This assertion, however,  
is very dependent on the Yukawa structure of the supersymmetric theory.
For instance, in the case of 
minimal supersymmetric SU(5) model \cite{susysu5}
the superpotential,
omitting SU(5) and flavor indices, is given by 
\begin{equation}
{\cal W}_{\rm SU(5)} = 
\frac{1}{4} \lambda_U \widehat \psi_{10} \widehat \psi_{10} \widehat  {\cal H}_{5} + 
\sqrt{2} \lambda_D \widehat \psi_{10} \widehat \psi_{\overline{5}} \widehat {\cal H}_{\overline{5}} +
\cdot \cdot \cdot
\end{equation}
where $\widehat \psi_{10}$ and $\widehat \psi_{\overline{5}}$ are matter chiral superfields
belonging to representations {\bf 10} and $\overline{{\bf 5}}$
of SU(5), respectively.
As in the supersymmetric generalization of the SM,
to generate fermion masses
we need two sets of Higgs superfields,  $\widehat {\cal H}_{5}$
and $\widehat {\cal H}_{\overline{5}}$, belonging to representations {\bf 5} and 
$\overline{{\bf 5}}$ of SU(5). In ordinary SUSY GUTs,
after integrating out the colored Higgs triplet,
the presence of Yukawa couplings in the superpotential leads directly to effective 
dimension-five interactions which omitting flavor indices are of the form,
\begin{eqnarray}
{\cal W}_{\rm dim \,5} & = & 
\frac{1}{M_{{\cal H}_{c}}} \left[\frac{1}{2}
\lambda_U  \lambda_{D} \left( Q Q\right) \left( Q L \right) + \right. \nonumber \\
&&\left. \lambda_U  \lambda_{D} \left( U E \right) \left( U D \right) \right],
\end{eqnarray}
where the operators $\left( Q Q\right) \left( Q L \right)$ and
$\left( U E \right) \left( U D \right) $ are totally antisymmetric in color indices.
Therefore flavor conservation in the superpotential would imply their cancellation,
\begin{eqnarray}
\left( Q Q \right) \left( Q L\right) &\equiv& 0, \\
\left( U E \right) \left( U D \right) &\equiv& 0.
\end{eqnarray}
In our case, we started assuming that
there is a symmetry that guarantees flavor conservation in the 
superpotential of the supersymmetric
unified theory, as is expected from any model generating flavor radiatively. 
Flavor violating Yukawa couplings are only generated at low energy after supersymmetry
breaking. We can see that after integrating out coloured Higgsses 
one can generate operators generically of the form,
 $$
 \frac{1}{M^{2}} ({\cal Z} {\cal Z}) QQQL.
 $$
These operators cannot generate directly dimension five operators 
because we assumed that only the auxiliary components
of the SUSY breaking ${\cal Z}$-fields acquire a vev, breaking the flavor symmetry.
Dimension five operators could be generated at higher orders.
Since tree level interactions with coloured Higgsinos are only possible
for the third family, the generation of a dimension five proton decay
operator would require two flavor mixing couplings between first
and third generation. On the other hand,
the Yukawa coupling of the form $({\bf Y}_{U})_{13}$
is first generated at two loops and is very suppressed,
as pointed out in Eq.~\ref{YUsup}.
As a consequence radiatively generated 
dimension five operators leading to proton decay 
are extremely suppressed in this scenario,
when compared with ordinary SUSY GUT predictions, which
generate flavor in the superpotential. 
Regarding the next dominant decay mode coming from dimension-six operators
via GUT gauge bosons. It has been shown that using the SuperKamiokande
limit, $\tau(p \rightarrow \pi^{0} e^{+})  > 5.3 \times 10^{33}$~years,
a lower bound on the heavy gauge boson mass, $M_{V}$, 
can be extracted, $M_{V} > 6.8 \times 10^{15}$~GeV. 
Furthermore, the proton decay rate for $M_{V} = M_{GUT}$ is far below
the detection limit that can be reached within the next years
\cite{Emmanuel-Costa:2003pu}. 
\section{CP violation}
The CP violation experimentally observed in the SM has not
been included in this analysis.
There is no doubt that CP violation
can be accommodated in this scenario since
the supersymmetry breaking sector provides numerous sources
of CP violation. 
If that is the case, 
we expect correlations between supersymmetric 
contributions to processes such as:
$K^{0}-\overline{K}^{0}$ mixing, 
electric dipole moments, $B \rightarrow \phi K_{S}$,
etc. 
The extension of this scenario to include CP-violation 
will be the subject of a future paper. 
\section*{Appendix}
For the calculation of the fermion mass ratios in the main text 
running quark masses were used.
These were calculated through scaling factors including 
QCD and QED renormalization effects, which 
can be determined using known solutions to the SM RGEs.
For the charged leptons our starting point are the well
known physical masses. 
For the top quark the starting point is the pole mass
from the PDG collaboration \cite{Hagiwara:fs}, 
\begin{eqnarray}
m_{t} &=& 174.3 \pm 5.1 ~\hbox{GeV}.
\end{eqnarray}
For the  bottom and charm quarks the running masses, 
$m_{b}(m_{b})_{\overline{MS}}$ 
and $m_{c}(m_{c})_{\overline{MS}}$ from Refs.~\cite{bottommass} \& 
\cite{Becirevic:2001yh} are used, 
\begin{eqnarray}
m_{b}(m_{b})_{\overline{MS}} &=& 4.25 \pm 0.25 ~\hbox{GeV},  \\
m_{c}(m_{c})_{\overline{MS}} &=& 1.26 \pm 0.05 ~\hbox{GeV}.
\end{eqnarray}
For the light quarks, u,d and s, the starting point is
the normalized $\overline{MS}$ values at $\mu=2$~GeV.
Original extractions \cite{Jamin:2001zr,Gamiz:2002nu} quoted in the literature 
have been rescaled as in \cite{Hagiwara:fs},
\begin{eqnarray}
m_{s}(2~\hbox{GeV})_{\overline{MS}} &=& 117 \pm 17 ~\hbox{MeV},  \\
m_{d}(2~\hbox{GeV})_{\overline{MS}} &=& 5.2 \pm 0.9~\hbox{MeV}, \\
m_{u}(2~\hbox{GeV})_{\overline{MS}} &=& 2.9 \pm 0.6 ~\hbox{MeV}. 
\end{eqnarray}
For completeness we also include here some functions
used in the main text. The $B_{0}$ function used in the calculation
of the one--loop finite corrections is given by
\begin{equation}
B_{0} (m_{1},m_{2}) = 1 +  \ln \left(\frac{{Q}^{2}}{m_{2}^{2}}\right)
+\frac{m_{1}^{2}}{m^{2}_{2}-m_{1}^{2}} 
 \ln \left(\frac{{m}^{2}_{2}}{m_{1}^{2}}\right).
\end{equation}
The following functions, extracted from Ref.~\cite{masieroFCNC},
are used in the calculation of $\Delta m_{K}$,
$\Delta m_{B}$ and $B(b\rightarrow s \gamma)$,
\begin{eqnarray}
f(x) &=& \frac{6(1 + 3x) \ln x + x^{3} - 9 x^{2} -9x +17}{6(x-1)^{5}},\\
g(x) &=& \frac{6x(1+x)\ln x -x^{3}  - 9x^{2} +9 x  + 1}{3(x-1)^{5}}, \\
M(x) &=& \frac{1 + 4x - 5x^{2} +4 x \ln x + 2 x^{2} \ln x}{2(1-x)^{4}}.
\end{eqnarray}
\acknowledgements
%
I am especially grateful to 
Xerxes Tata and Carlos Wagner
for their insightful suggestions.
I thank to Arcadi Santamaria, whose original 
question triggered my interest in this topic. 
I thank to Alex Kagan for his comments regarding the calculation of
$B(b\rightarrow s\gamma)$.
I thank Radovan Dermisek for his suggestions regarding the 
higher order generation of Yukawa couplings and proton decay operators. 
I am also grateful to Ernest Ma, Tao Han, 
Liliana Velasco-Sevilla, C.P.~Yuan and Naoyuki Haba for interesting comments. 
I would like to thank, for their hospitality
during the elaboration of this paper, 
to the physics departments of the University of Madison, 
University of Michigan, Michigan State University, 
Argonne National Laboratory, University of California at Irvine
and University of California at Riverside.  
Finally I would like to thank H.~Guler for many suggestions.
This work is supported by the DOE grant number DE-FG03-94ER40833.



\begin{references}

\bibitem{Ferrandis:2002ws}
J.~Ferrandis,
Phys.\ Rev.\ D {\bf 68}, 015001 (2003)


\bibitem{Weinberg:1971nd}
S.~Weinberg,
Phys.\ Rev.\ D {\bf 5}, 1962 (1972);

\bibitem{Weinberg:1972ws}
S.~Weinberg,
Phys.\ Rev.\ Lett.\  {\bf 29} (1972) 388.

\bibitem{Ibanez:1982xg}
L.~E.~Ibanez,
Phys.\ Lett.\ B {\bf 117}, 403 (1982);

\bibitem{wyler}
W.~Buchmuller and D.~Wyler,
Phys.\ Lett.\ B {\bf 121}, 321 (1983);
A.~B.~Lahanas and D.~Wyler,
Phys.\ Lett.\ B {\bf 122}, 258 (1983);

\bibitem{Hall:1985dx}
L.~J.~Hall, V.~A.~Kostelecky and S.~Raby,
Nucl.\ Phys.\ B {\bf 267}, 415 (1986).

\bibitem{Banks:1987iu}
T.~Banks,
Nucl.\ Phys.\ B {\bf 303}, 172 (1988);

\bibitem{Kagan:1987wf}
A.~L.~Kagan and C.~H.~Albright,
Phys.\ Rev.\ D {\bf 38}, 917 (1988).

\bibitem{Ma:1988fp}
E.~Ma,
Phys.\ Rev.\ D {\bf 39}, 1922 (1989);

\bibitem{softRadSusy}
E.~Ma and D.~Ng,
Phys.\ Rev.\ Lett.\  {\bf 65}, 2499 (1990);
E.~Ma and K.~McIlhany,
Mod.\ Phys.\ Lett.\ A {\bf 6}, 1089 (1991).
R.~Hempfling,
Phys.\ Rev.\ D {\bf 49}, 6168 (1994);
C.~Hamzaoui and M.~Pospelov,
Eur.\ Phys.\ J.\ C {\bf 8}, 151 (1999);
K.~S.~Babu, B.~Dutta and R.~N.~Mohapatra,
Phys.\ Rev.\ D {\bf 60}, 095004 (1999);
J.~M.~Frere and E.~Ma,
Phys.\ Rev.\ D {\bf 68}, 051701 (2003)

\bibitem{Borzumati:1999sp}
F.~Borzumati, G.~R.~Farrar, N.~Polonsky and S.~Thomas,
Nucl.\ Phys.\ B {\bf 555}, 53 (1999);
F.~M.~Borzumati, G.~R.~Farrar, N.~Polonsky and S.~Thomas,
[arXiv:hep-ph/9805314].

\bibitem{kagan}
B.~S.~Balakrishna, A.~L.~Kagan and R.~N.~Mohapatra,
Phys.\ Lett.\ B {\bf 205}, 345 (1988);
A.~L.~Kagan,
Phys.\ Rev.\ D {\bf 40}, 173 (1989).


\bibitem{Arkani-Hamed:1995fq}
N.~Arkani-Hamed, H.~C.~Cheng and L.~J.~Hall,
Nucl.\ Phys.\ B {\bf 472}, 95 (1996);
N.~Arkani-Hamed, H.~C.~Cheng and L.~J.~Hall,
Phys.\ Rev.\ D {\bf 54}, 2242 (1996);

\bibitem{su5relations}
N.~V.~Krasnikov,
Phys.\ Lett.\ B {\bf 302} (1993) 59;
J.~L.~Diaz-Cruz, H.~Murayama and A.~Pierce,
Phys.\ Rev.\ D {\bf 65} (2002) 075011;

\bibitem{habajavi}
J.~Ferrandis and N.~Haba,
''Supersymmetry breaking as the origin of flavor'',
[arXiv:hep-ph/0404077]

\bibitem{Barbieri:1996ww}
R.~Barbieri, L.~J.~Hall, S.~Raby and A.~Romanino,
Nucl.\ Phys.\ B {\bf 493}, 3 (1997)

\bibitem{fritzsch}
H.~Fritzsch,
Nucl.\ Phys.\ B {\bf 155}, 189 (1979);
H.~Fritzsch and Z.~z.~Xing,
Prog.\ Part.\ Nucl.\ Phys.\  {\bf 45}, 1 (2000)

\bibitem{radovan}
I thank R.~Dermisek for pointing me out this possibility.

\bibitem{susyFC}
J.~R.~Ellis and D.~V.~Nanopoulos,
Phys.\ Lett.\ B {\bf 110}, 44 (1982);
R.~Barbieri and R.~Gatto,
Phys.\ Lett.\ B {\bf 110}, 211 (1982);
M.~Suzuki,
Phys.\ Lett.\ B {\bf 115}, 40 (1982);
T.~Inami and C.~S.~Lim,
Nucl.\ Phys.\ B {\bf 207}, 533 (1982);
A.~B.~Lahanas and D.~V.~Nanopoulos,
Phys.\ Lett.\ B {\bf 129}, 461 (1983).

\bibitem{Kagan:1994qg}
A.~L.~Kagan,
Phys.\ Rev.\ D {\bf 51}, 6196 (1995).

\bibitem{masieroFCNC}
F.~Gabbiani and A.~Masiero,
Nucl.\ Phys.\ B {\bf 322}, 235 (1989);
S.~Bertolini, F.~Borzumati, A.~Masiero and G.~Ridolfi,
Nucl.\ Phys.\ B {\bf 353}, 591 (1991);
F.~Gabbiani, E.~Gabrielli, A.~Masiero and L.~Silvestrini,
Nucl.\ Phys.\ B {\bf 477}, 321 (1996);

\bibitem{Hagiwara:fs}
K.~Hagiwara {\it et al.}  [Particle Data Group Collaboration],
Phys.\ Rev.\ D {\bf 66}, 010001 (2002) 
and 2003 off-year partial update for the 2004 edition available on the PDG WWW pages (URL: http://pdg.lbl.gov/)

\bibitem{Bertolini:1986tg}
S.~Bertolini, F.~Borzumati and A.~Masiero,
Phys.\ Lett.\ B {\bf 192}, 437 (1987).

\bibitem{bottommass} 
H.~Baer, J.~Ferrandis, K.~Melnikov and X.~Tata,
Phys.\ Rev.\ D {\bf 66}, 074007 (2002);
K. Melnikov and A. Yelkhovsky, Phys. Rev. {\bf D59}, 114009 (1999);
A.H. Hoang, Phys. Rev. {\bf D61}, 034005 (2000); 
M. Beneke and A. Signer, Phys. Lett. {\bf B471}, 233 (1999);
A.~A.~Penin and A.~A.~Pivovarov, 
Nucl.\ Phys. {\bf B549}, 217 (1999);
G.~Rodrigo, A.~Santamaria and M.~S.~Bilenky,
Phys.\ Rev.\ Lett.\  {\bf 79} (1997) 193


\bibitem{Becirevic:2001yh}
D.~Becirevic, V.~Lubicz and G.~Martinelli,
Phys.\ Lett.\ B {\bf 524}, 115 (2002)

\bibitem{Jamin:2001zr}
M.~Jamin, J.~A.~Oller and A.~Pich,
Eur.\ Phys.\ J.\ C {\bf 24}, 237 (2002)

\bibitem{Gamiz:2002nu}
E.~Gamiz, M.~Jamin, A.~Pich, J.~Prades and F.~Schwab,
JHEP {\bf 0301}, 060 (2003)


\bibitem{susysu5}
E.~Witten,
Nucl.\ Phys.\ B {\bf 188}, 513 (1981);
N.~Sakai,
Z.\ Phys.\ C {\bf 11}, 153 (1981);
S.~Dimopoulos and H.~Georgi,
Nucl.\ Phys.\ B {\bf 193}, 150 (1981).

\bibitem{Emmanuel-Costa:2003pu}
D.~Emmanuel-Costa and S.~Wiesenfeldt,
Nucl.\ Phys.\ B {\bf 661}, 62 (2003)

\end{references}
\end{document}